\documentclass[preprint,showpacs,preprintnumbers,amsmath,amssymb]{revtex4}

\pdfoutput=1

\usepackage{graphicx}
\usepackage{dcolumn}
\usepackage{bm}
\usepackage{amssymb}

\begin{document}
\title{
Stochastic resonance in bistable systems with 
nonlinear dissipation and multiplicative noise: 
A microscopic approach
}

\author{Hideo Hasegawa}
\altaffiliation{hideohasegawa@goo.jp}
\affiliation{Department of Physics, Tokyo Gakugei University,  
Koganei, Tokyo 184-8501, Japan}%

\date{\today}
\begin{abstract}
The stochastic resonance (SR) in bistable systems has been extensively discussed
with the use of {\it phenomenological} Langevin models.
By using the {\it microscopic}, generalized Caldeira-Leggett (CL) model,
we study in this paper, SR of an open bistable system 
coupled to a bath with a nonlinear system-bath interaction.
The adopted CL model yields the non-Markovian Langevin equation
with nonlinear dissipation and state-dependent diffusion
which preserve the fluctuation-dissipation relation (FDR).
From numerical calculations, we find the following: 
(1) the spectral power amplification (SPA) exhibits SR not only for $a$ and $b$ but also for $\tau$ 
while the stationary probability distribution function is independent of them
where $a$ ($b$) denotes the magnitude of multiplicative (additive) noise 
and $\tau$ expresses the relaxation time of colored noise; 
(2) the SPA for coexisting additive and multiplicative noises
has a single-peak but two-peak structure as functions of $a$, $b$ and/or $\tau$.
These results (1) and (2) are qualitatively different from previous ones obtained by 
phenomenological Langevin models where the FDR is indefinite or not held.
\end{abstract}

        

\maketitle
\newpage
\section{Introduction}
In the last three decades since a seminal work of Benzi {\it et. al.} \cite{Benzi81},
extensive studies have been made on the stochastic resonance (SR)
(for a recent review on SR, see Refs. \cite{Gamma98,Lindner04}).
Theoretical studies on SR were made initially with the use of 
the overdamped Langevin model subjected to white noise.
It has been shown that the signal-to-noise ratio (SNR) of nonlinear systems to 
a small periodic signal is resonantly enhanced under some optimum level of noise, 
whose effect is called SR \cite{Gamma98,Lindner04}.
The resonant activation occurs when the frequency of an applied periodic signal
is near the Kramer escape rate of the transition 
from one potential minimum to another. 
Subsequent studies on SR examined effects of colored noise
(for a review on colored noise, see Ref. \cite{Hanggi95}). 
Calculations with the use of the universal colored-noise approximation (UCNA) \cite{Jung87}
have shown that noise color suppresses SR monotonically 
with increasing its relaxation time $\tau$ \cite{Hanggi84,Hanggi93,Gamma89}.
Ref. \cite{Neiman96} has claimed that when $\tau$ is increased, the SR is first
suppressed for small $\tau$ but enhanced for large $\tau$ 
with the minimum at intermediate $\tau$.
In recent years, studies on SR have been made by using more sophisticated noise:
correlated additive white noise and multiplicative colored noise
\cite{Li95,Fu99,Jia00,Jia01,Luo03}. 
SNR of the Langevin model
subjected to such noises is shown to exhibit a complicated behavior 
as functions of magnitudes of additive and multiplicative noises
and the correlation strength between the two noises \cite{Jia00,Jia01,Luo03}.
For example, SNR may have the bimodal structure when these parameters are varied.

It should be noted that Langevin equations
having been adopted for studies on SR are phenomenological ones which have 
some deficits such that a lack of the fluctuation-dissipation relation (FDR), 
related discussion being given in Sec. IV.
We expect that in almost all real, non-isolated systems, the environment acts 
as a thermal bath which is a source of noise.
The microscopic origin of additive noise has been proposed
within a framework of Caldeira-Leggett (CL) system-bath Hamiltonian
\cite{Ford65,Ullersma66,Caldeira81}.
The CL model with a linear system-bath coupling leads to the non-Markovian 
Langevin equation  where memory kernel and noise term are expressed in terms of
bath variables and they satisfy the FDR \cite{Ford65,Ullersma66,Caldeira81}.
The FDR implies that dissipation and diffusion come from the same origin.
By using the generalized CL model including
a nonlinear system-bath coupling, we may obtain the non-Markovian
Langevin equation with multiplicative noise which preserves the FDR 
\cite{Lindenberg81,Pollak93}.
The nature of nonlinear dissipation and multiplicative noise
has been recently explored with renewed interest 
\cite{Barik05,Plyukhin07,Farias09}.
Nonlinear dissipation and multiplicative noise
have been recognized as important ingredients in several fields
such as mesoscopic scale systems \cite{Zaitsev09,Eichler11} 
and ratchet problems \cite{Magnasco93,Julicher97,Reimann02,Porto00}.

Quite recently, we have studied the generalized CL model for a harmonic oscillator which includes
nonlinear nonlocal dissipation and multiplicative diffusion \cite{Hasegawa11d}.
It has been shown that the stationary probability distribution function (PDF) 
is independent of magnitudes of additive and multiplicative noises and 
of the relaxation time of colored noise \cite{Barik05,Plyukhin07,Farias09,Hasegawa11d}, 
although the response to applied input depends on noise parameters \cite{Hasegawa11d}. 
This is in contrast with previous studies 
\cite{Sakaguchi01,Anteneodo03,Hasegawa07} which show that
the stationary PDF of the Langevin model for the harmonic potential
is Gaussian or non-Gaussian, depending on magnitudes of additive and multiplicative noises.
We expect that when the generalized CL model with a nonlinear coupling 
\cite{Barik05,Plyukhin07,Farias09,Hasegawa11d} 
is applied to a bistable system, properties of the calculated SR may be rather different
from those having been derived by the phenomenological Langevin models 
\cite{Hanggi84,Hanggi93,Hanggi95,Gamma89,Neiman96,Li95,Fu99,Jia00,Jia01,Luo03}.
It is the purpose of the present paper to make a detailed study on SR
of an open bistable system described by the microscopic, generalized CL model 
with nonlinear nonlocal dissipation and state-dependent diffusion \cite{Hasegawa11d}.
Although the UCNA has been employed in many studies  
on SR for colored noise \cite{Li95,Fu99,Jia00,Jia01,Luo03}
because it is exact both for $\tau=0.0$ and $\tau=\infty$ \cite{Jung87},
its validity within $O(\tau)$ is not justified \cite{Timmermann01,Hasegawa07b}.
In this study, we investigate SR for a wide range of $\tau$ value
by simulations. 

The paper is organized as follows.
In Sec. II, we briefly summarize the non-Markovian Langevin equation 
derived from the generalized CL model with OU colored noise 
\cite{Barik05,Plyukhin07,Farias09,Hasegawa11d}.
In Sec. III, after studying the stationary PDF, we investigate SR of an open bistable 
system for three cases: additive noise, multiplicative noise, 
and coexisting additive and multiplicative noises. In Sec. IV, we examine
a validity of the Markovian Langevin equations with the local approximation,
which have been widely employed in previous studies on SR \cite{Hanggi84,Hanggi93,Hanggi95,Gamma89}.
Sec. V is devoted to our conclusion.

\section{The generalized CL model}
\subsection{Non-Markovian Langevin equation}

We consider a system of a Brownian particle coupled to 
a bath consisting of $N$-body uncoupled oscillators, which is described by
the generalized CL model \cite{Barik05,Plyukhin07,Farias09,Hasegawa11d},
\begin{eqnarray}
H &=& H_S + H_B + H_I,
\label{eq:A1}
\end{eqnarray}
with
\begin{eqnarray}
H_S &=& \frac{p^2}{2}+ V(x) - x f(t), 
\label{eq:A1b} \\
H_B+H_I &=& \sum_{n=1}^N \left\{ \frac{p_n^2}{2 m_n}
+ \frac{m_n \omega_n^2}{2}\left( q_n -\frac{c_n \phi(x)}{m_n \omega_n^2}\right)^2 \right\}.
\label{eq:A1c} 
\end{eqnarray}
Here $H_S$, $H_B$ and $H_I$ express Hamiltonians of the system, bath and interaction, 
respectively; $x$, $p$ and $V(x)$ denote position, momentum and potential, respectively, 
of the system; $q_n$, $p_n$, $m_n$ and $\omega_n$ stand for position, momentum, mass 
and frequency, respectively, of bath; the system couples to the bath nonlinearly 
through a function $\phi(x)$; 
$f(t)$ expresses an applied external force.
The original CL model adopts a linear system-bath coupling
with $\phi(x)=x$ in Eq. (\ref{eq:A1c}) which yields additive noise \cite{Caldeira81}.
By using the standard procedure, we obtain the generalized Langevin 
equation given by \cite{Caldeira81,Lindenberg81,Pollak93,Barik05,Hasegawa11d}
\begin{eqnarray}
\ddot{x}(t) &=& -V'(x(t))
-\phi'(x(t)) \int_0^t \gamma(t-t') \:\phi'(x(t')) \:\dot{x}(t')\:dt'
+\phi'(x(t))\:\zeta(t)+f(t),
\label{eq:A4}
\end{eqnarray}
with
\begin{eqnarray}
\gamma(t) &=& \sum_{n=1}^N \left(\frac{c_n^2}{m_n \omega_n^2} \right) \cos \omega_n t, 
\label{eq:A5}\\
\zeta(t) 
&=& \sum_{n=1}^N \left\{ \left[\frac{m_n \omega_n^2}{c_n}\: q_n(0) 
-\phi(x(0)) \right]\left( \frac{c_n^2}{m_n\omega_n^2} \right) \cos \omega_n t
+\left( \frac{c_n p_n(0)}{m_n \omega_n} \right) \sin \omega_n t \right\}, 
\label{eq:A6}
\end{eqnarray}
where $\gamma(t-t')$ denotes the non-local memory kernel and $\zeta(t)$ stands for noise,
and dot and prime stand for derivatives with respect
to time and argument, respectively.
Dissipation and diffusion terms given by Eqs. (\ref{eq:A5}) and (\ref{eq:A6}),
respectively, satisfy the FDR,
\begin{eqnarray}
\left< \zeta(t) \zeta(t') \right>_0
&=& k_B T \: \gamma(t-t'),
\label{eq:A7}
\end{eqnarray}
where the bracket $\langle \cdot \rangle_0$ stands for the average over initial states 
of $q_n(0)$ and $p_n(0)$ \cite{Lindenberg81,Pollak93,Barik05}.

We adopt the OU process for the kernel $\gamma(t-t')$ given by
\begin{eqnarray}
\gamma(t-t') &=& \left( \frac{\gamma_0}{\tau} \right)
e^{-\vert t-t' \vert/\tau},
\label{eq:A8}
\end{eqnarray}
where $\gamma_0$ and $\tau$ stand for the strength and relaxation time,
respectively, of colored noise.
The OU colored noise may be generated by the differential equation,
\begin{eqnarray}
\dot{\zeta}(t) &=& -\frac{\zeta(t)}{\tau}
+\frac{\sqrt{2 k_B T \gamma_0} }{\tau} \: \xi(t),
\label{eq:A14b}
\end{eqnarray}
where $\xi(t)$ expresses white noise with
\begin{eqnarray}
\left< \xi(t) \right> &=& 0, \;\;\;\;
\left< \xi(t) \xi(t')\right> = \delta(t-t').
\label{eq:A15}
\end{eqnarray}
Equations (\ref{eq:A14b}) and (\ref{eq:A15}) lead to the PDF
and correlation of colored noise given by
\begin{eqnarray}
P(\zeta) &\propto& e^{-(\beta \tau/2 \gamma_0) \:\zeta^2},
\label{eq:A17} \\
\left< \zeta(t) \zeta(t')\right> 
&=& \left( \frac{k_B T \gamma_0}{\tau}\right) \:e^{-\vert t-t'\vert/\tau}
= k_B T \:\gamma(t-t'), 
\label{eq:A16}
\end{eqnarray}
where $\beta=1/k_B T$.

In order to transform the non-Markovian Langevin equation given by Eq. (\ref{eq:A4}) 
into more tractable multiple differential equations,
we introduce a new variable $u(t)$ \cite{Pollak93,Farias09},
\begin{eqnarray}
u(t) &=& - \int_0^t \gamma(t-t') \phi'(x(t')) \dot{x}(t')\:dt',
\label{eq:A10}
\end{eqnarray}
which yields four first-order differential equations for $x(t)$, $p(t)$, $u(t)$ 
and $\zeta(t)$ given by
\begin{eqnarray}
\dot{x}(t) &=& p(t), 
\label{eq:A11}\\
\dot{p}(t) &=& -V'(x)+\phi'(x(t)) \:u(t) +f(t)
+ \phi'(x(t)) \:\zeta(t), 
\label{eq:A12}\\
\dot{u}(t) &=& - \frac{u(t)}{\tau} 
- \left( \frac{\gamma_0}{\tau} \right) \phi'(x(t)) \:p(t), 
\label{eq:A13}\\
\dot{\zeta}(t) &=& -\frac{\zeta(t)}{\tau}
+\frac{\sqrt{2 k_B T \gamma_0} }{\tau} \: \xi(t).
\label{eq:A14}
\end{eqnarray}
From Eqs. (\ref{eq:A11})-(\ref{eq:A14}), we obtain the multi-variate 
Fokker-Planck equation (FPE) for distribution of $P(x,p,u,\zeta,t)$,
\begin{eqnarray}
\frac{\partial P(x,p,u,\zeta,t)}{\partial t} 
&=& - \frac{\partial}{\partial x} \:p \: P(x,p,u,\zeta,t) \nonumber \\
&-& \frac{\partial}{\partial p}\left[-V'(x)+f(t)+\phi'(x) u+\phi'(x)\zeta
\right]P(x,p,u,\zeta,t) \nonumber \\
&-& \frac{\partial}{\partial u}\left[ \frac{u}{\tau}
+\left( \frac{\gamma_0}{\tau} \right) \:\phi'(x) p \right] P(x,p,u,\zeta,t)
+ \frac{\partial}{\partial \zeta} \left( \frac{\zeta}{\tau} \right)
P(x,p,u,\zeta,t) \nonumber \\
&+& \left( \frac{k_B T \gamma_0}{\tau^2} \right) \frac{\partial^2}{\partial \zeta^2} 
P(x,p,u,\zeta,t). 
\label{eq:A18}
\end{eqnarray}
The stationary PDF with $f(t)=0$ is given by \cite{Pollak93}
\begin{eqnarray}
P(x, p) &=& \int P(x, p, u, \zeta) \;du \;d\zeta
\propto e^{-\beta [p^2/2+V(x)]}.
\label{eq:Y8}
\end{eqnarray}

\subsection{Markovian Langevin equation}
It is worthwhile to examine the local limit of Eq. (\ref{eq:A7}),
\begin{eqnarray}
\left< \zeta(t) \zeta(t')\right>_0 &=& k_B T \:\gamma(t-t') 
= 2 k_B T \:\gamma_0 \:\delta(t-t'),
\label{eq:B1}
\end{eqnarray}
which leads to the Markovian Langevin equation given by
\begin{eqnarray}
\ddot{x}(t) &=& -V'(x(t)) -\gamma_0 \phi'(x(t))^2 \:\dot{x}(t)
+\sqrt{2 k_B T} \: \phi'(x)  \:\xi(t) +f(t),
\label{eq:B2}
\end{eqnarray}
preserving the FDR.
It is evident that the Markovian Langevin equation becomes a good approximation
of the non-Markovian one in the limit of $\tau \rightarrow 0$.

From Eq. (\ref{eq:B2}), we obtain first-order differential equations,
\begin{eqnarray}
\dot{x}(t) &=& p(t), \\
\dot{p}(t) &=& -V'(x)-\gamma_0 \phi'(x(t))^2 p(t)
+\sqrt{2 k_B T \gamma_0} \;\phi'(x(t)) \: \xi(t) +f(t).
\end{eqnarray}
The relevant FPE for the PDF of $P(x,p,t)$ is expressed by
\begin{eqnarray}
\frac{\partial P(x,p,t)}{\partial t} 
&=& - \frac{\partial }{\partial x} \:p \:P(x,p,t) 
+ \frac{\partial}{\partial p}
\left[ V'(x)-f(t)+\gamma_0 \phi'(x)^2 \:p \right] P(x,p,t)
\nonumber  \\
&+& k_B T \:\gamma_0 
\;\frac{\partial}{\partial p} \phi'(x) 
\frac{\partial}{\partial p} \phi'(x) P(x,p,t).
\label{eq:B3}
\end{eqnarray}
The stationary distribution of Eq. (\ref{eq:B3}) with $f(t)=0$ is given by
\begin{eqnarray}
P(x, p) &\propto &  \:e^{-\beta[p^2/2+V(x)]},
\label{eq:B4}
\end{eqnarray}
which is consistent with the result of the non-Markovian Langevin
equation given by Eq. (\ref{eq:Y8}).

\subsection{Overdamped Langevin equation}

We will derive the overdamped Langevin equation from the Markovian
Langevin model given by Eq. (\ref{eq:B2}).
The overdamped limit of the Markovian Langevin equation
is conventionally derived with setting $\ddot{x}=0$ in Eq. (\ref{eq:B2}).
This is, however, not the case because dissipation and diffusion terms
are state dependent in Eq. (\ref{eq:B2}).
Sancho, San Miguel and D\"{u}rr \cite{Sancho82} have developed 
an adiabatic elimination procedure to obtain an exact overdamped Langevin equation and its FPE.
In order to adopt their method \cite{Sancho82}, we rewrite Eq. (\ref{eq:B2}) as
\begin{eqnarray}
\ddot{x}(t) 
&=& -V'(x(t)) -\lambda(x(t)) \dot{x}(t)
+g(x(t))  \:\xi(t)+f(t),
\label{eq:D1}
\end{eqnarray}
with
\begin{eqnarray}
\lambda(x) &=& \gamma_0 \phi'(x)^2, 
\label{eq:D2}\\
g(x) &=& \sqrt{2 k_B T \gamma_0} \:\phi'(x).
\label{eq:D3}
\end{eqnarray}
By the adiabatic elimination in Eq. (\ref{eq:D1}) after Ref. \cite{Sancho82}, the FPE 
in the Stratonovich interpretation is given by 
\begin{eqnarray}
\frac{\partial P(x,t)}{\partial t} 
&=& \frac{\partial}{\partial x} \frac{1}{\lambda(x)}
\left[ V'(x)-f(t) +k_B T \frac{\partial }{\partial x} \right] P(x,t),
\label{eq:D4}
\end{eqnarray}
where we employ the relation: $g(x)^2= 2 k_B T \lambda(x)$ derived from 
Eqs. (\ref{eq:D2}) and (\ref{eq:D3}).
The relevant Langevin equation is given by
\cite{Sancho82} 
\begin{eqnarray}
\dot{x} &=& - \frac{[V'(x)-f(t)]}{\lambda(x)}
-\frac{1}{2 \lambda(x)^2} \:g'(x) g(x) +\frac{g(x)}{\lambda(x)} \:\xi(t).
\label{eq:D5}
\end{eqnarray}
Note that the second term of Eq. (\ref{eq:D5}) does not appear when we obtain the
overdamped Langevin equation by simply setting $\ddot{x}=0$ in Eq. (\ref{eq:D1}). 
It is easy to see that the stationary distribution of Eq. (\ref{eq:D4}) 
with $f(t)=0$ is given by
\begin{eqnarray}
P(x) &\propto& e^{-\beta V(x)}.
\label{eq:D6}
\end{eqnarray}

\section{Bistable systems}
\subsection{Stationary PDF}
We apply the generalized CL model mentioned in the preceding section to a bistable system,
where $V(x)$ and $\phi(x)$ in Eqs. (\ref{eq:A1})-(\ref{eq:A1c}) are given by
\begin{eqnarray}
V(x) &=& \frac{x^4}{4}-\frac{x^2}{2}, 
\label{eq:E1}\\
\phi(x) &=& \frac{a x^2}{2} + b x
\hspace{1cm}\mbox{($a \geq 0, \;b \geq 0$)},
\label{eq:E2}
\end{eqnarray}
$a$ and $b$ denoting magnitudes of multiplicative and additive noises, respectively.
The symmetric bistable potential given by Eq. (\ref{eq:E1}) has two stable minima 
at $x= \pm 1.0$ and unstable maximum at $x=0.0$. The height of the potential barrier 
is given by $\Delta \equiv V(0)-V(\pm 1)=1/4$.

First we show calculations of the stationary PDF.
We have solved Eqs. (\ref{eq:A11})-(\ref{eq:A14}) by using
the Heun method \cite{Note1,Note2} with a time step of 0.001 for parameters of 
$\gamma_0=1.0$ and $k_B T=\Delta=0.25$.
Simulations of Eqs. (\ref{eq:A11})-(\ref{eq:A14}) have been performed
for $0 \leq t < 1000.0$ with discarding initial results at $t < 200.0$.
Calculated results to be reported are averaged over $10\;000$ runs with initial states of
Gaussian-distributed $x(0)$, $p(0)$ and $\zeta(0)$ with 
$\langle x(0) \rangle=\langle p(0) \rangle =\langle \zeta(0) \rangle = 0$,
$\langle x(0)^2 \rangle=\langle p(0) ^2 \rangle=k_B T$,
and $\langle \zeta(0)^2 \rangle=k_B T \gamma_0/\tau$.
An initial condition of $u(0)=0$ has been used in all simulations.

\begin{figure}
\begin{center}
\includegraphics[keepaspectratio=true,width=80mm]{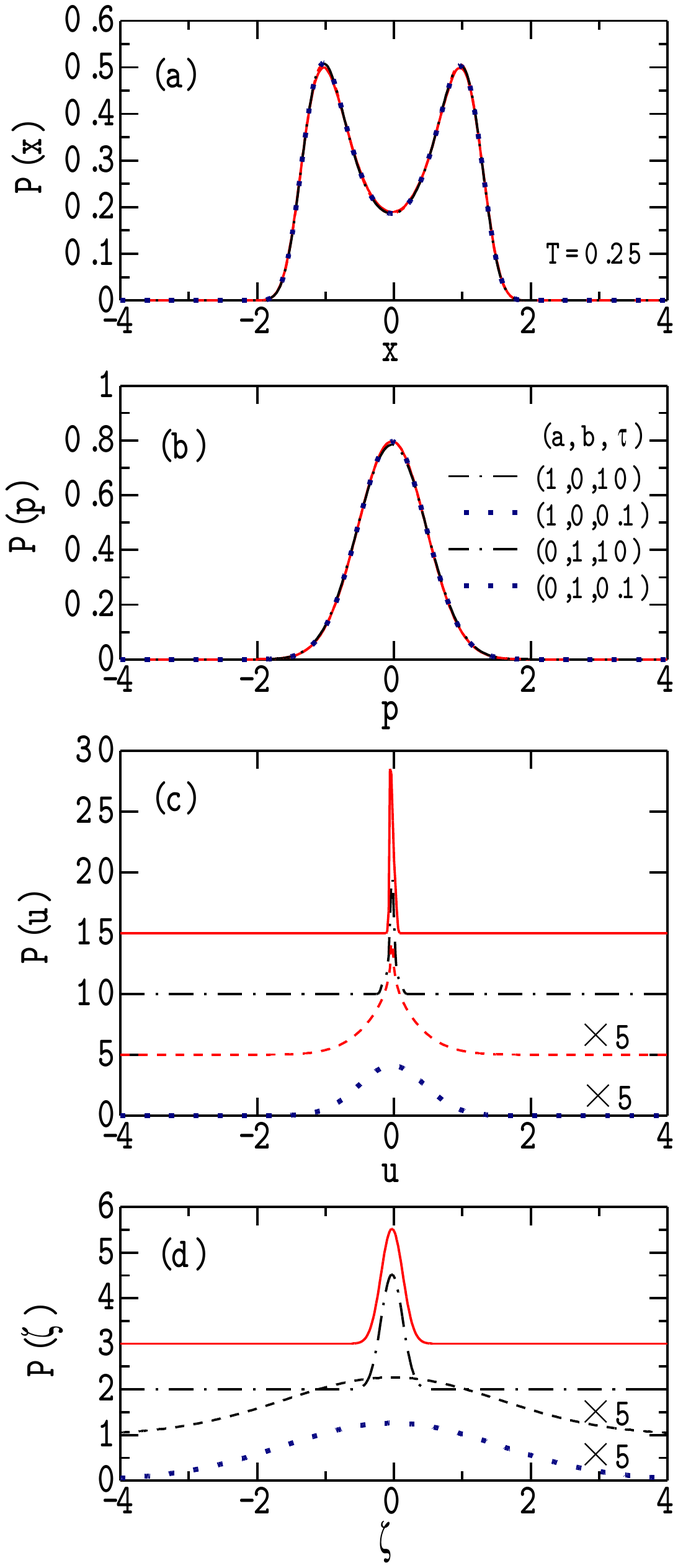}
\end{center}
\caption{
(Color online) 
Marginal PDFs of (a) $P(x)$, (b) $P(p)$, (c) $P(u)$ and (d) $P(\zeta)$
for $(a,b,\tau)=(1,0,10.0)$ (solid curves), $(1,0,0.1)$ (dashed curves),
$(0,1,10.0)$ (chain curves), and $(0,1,0.1)$ (dotted curves) calculated 
by simulations with $\gamma_0=1.0$ and $k_B T=\Delta=0.25$.
$P(x)$ and $P(p)$ in (a) and (b) are indistinguishable 
among the four sets of $(a, b, \tau)$.
$P(u)$ and $P(\zeta)$ for $(a,b,\tau)=(1,0,0.1)$ and (0,1,0.1) are
multiplied by a factor of five in (c) and (d) 
where PDFs are arbitrarily shifted for a clarity of figures.
}
\label{fig1}
\end{figure}
 
Calculated marginal PDFs of $P(x)$, $P(p)$, $P(u)$ and $P(\zeta)$ 
for $f(t)=0$ are shown in Figs. \ref{fig1}(a), (b), (c) and (d), respectively,
where four sets of parameters are adopted: 
$(a,b,\tau)=(1, 0,10.0)$ (solid curves), $(1,0,0.1)$ (dashed curves),
$(0,1,10.0)$ (chain curves) and $(0,1,0.1)$ (dotted curves).
We note that $P(x)$ and $P(p)$ in Figs. \ref{fig1}(a) and (b) are independent of
the parameters of $(a, b, \tau)$ although $P(u)$ and $P(\zeta)$ 
in Figs. \ref{fig1}(c) and (d) depend on them.
Calculated $P(x)$ and $P(p)$ are in good agreement with marginal PDFs obtained by
\begin{eqnarray}
P(x) &=& \int P(x,p)\; dp \propto e^{-\beta V(x)}, \\
P(p) &=& \int P(x,p)\; dx \propto e^{-\beta p^2/2},
\end{eqnarray}
which show that $P(x)$ has the typical bimodal structure 
while $P(p)$ is Gaussian.
Equation (\ref{eq:A17}) shows that $P(\zeta)$ is the Gaussian PDF whose variance
depends on $\tau$ for fixed $\gamma_0$ and $T$.
In contrast, Fig. \ref{fig1}(c) shows that $P(u)$ is Gaussian PDF
for additive noise  but non-Gaussian PDF for multiplicative noise.
Properties of stationary PDFs for bistable potential shown in Fig. 1 are similar to
those for harmonic oscillators except for $P(x)$ (see Fig. 1 of Ref. \cite{Hasegawa11d}). 

\begin{figure}
\begin{center}
\includegraphics[keepaspectratio=true,width=100mm]{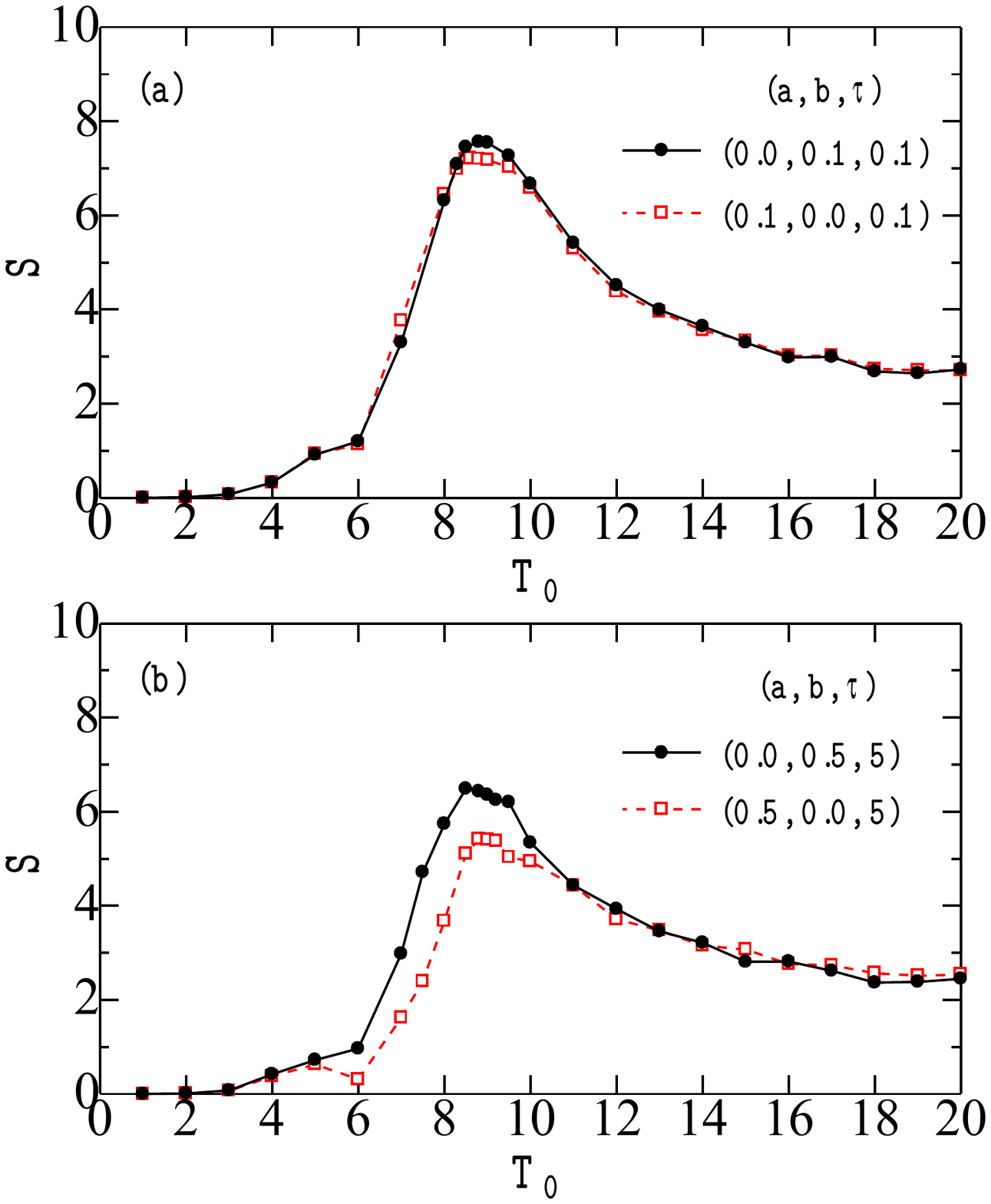}
\end{center}
\caption{
(Color online) 
(a) SPA ($S$) as a function of a period of an input ($T_0$)
for $(a, b, \tau)=(0.0, 0.1, 0.1)$ (solid curve) and $(0.1, 0.0, 0.1)$ (dashed curve).
(b) $S$ as a function of $T_0$ for $(a, b, \tau)=(0.0, 0.5, 5.0)$ (solid curve)
and $(0.5, 0.0, 5.0)$ (dashed curve).
}
\label{fig2}
\end{figure}

\subsection{Stochastic resonance}
\subsubsection{Spectral power amplification}
We investigate SR, applying a sinusoidal input given by
\begin{eqnarray}
f(t) &=& g \:\cos\left(\frac{2 \pi t}{T_0} \right) =g\:\cos\omega_0 t,
\label{eq:C1}
\end{eqnarray}
where $g$, $T_0$ and $\omega_0$ ($=2 \pi/T_0$) 
denote its magnitude, period and frequency, respectively.
A sinusoidal input given by Eq. (\ref{eq:C1}) yields 
an averaged output given by
\begin{eqnarray}
\mu(t) &=& \langle x(t) \rangle, 
\label{eq:C2}
\end{eqnarray}
where $\langle \cdot \rangle$ denotes the average over trials.
We evaluate the spectral power amplification (SPA) $S$ defined by
\begin{eqnarray}
S &=& \frac{\vert \: \mu[\omega_0] \:\vert^2}{\vert \:f[\omega_0] \:\vert^2},
\end{eqnarray}
where $f[\omega]$ and $\mu[\omega]$ are Fourier transformations of $f(t)$
and $\mu(t)$, respectively. 
Simulation procedures are the same as those for a calculation of stationary PDFs
except for initial conditions of $x(0)$ and $p(0)$ which are taken 
to be $x(0)=1.0$ and $p(0)=0.0$. 
Direct simulations for Eqs. (\ref{eq:A11})-(\ref{eq:A14})
have been made for sets of $a$, $b$, $\tau$, $g$ and $T_0$ averaged over $10 \;000$ runs.

Figure \ref{fig2}(a) shows SPA ($S$) as a function of a period of an external input
($T_0$) with $g=0.05$ for two sets of parameters: 
$(a, b, \tau)=(0.0, 0.1, 0.1)$ (solid curve) and $(0.1, 0.0, 0.1)$ (dashed curve), 
the former (latter) corresponding to additive (multiplicative) noise only.
SPAs for both sets of parameters have maxima at $T_0 \sim 9$.
It is interesting that $S$ for additive noise nearly coincides with 
that for multiplicative noise. 
This coincidence, however, is not realized for a larger $\tau$ and/or $b$
as shown in Fig. \ref{fig2}(b) where similar plots of the $T_0$ dependence of $S$ are presented
for different two sets of parameters: $(a, b, \tau)=(0.0, 0.5, 5)$ (solid curve) 
and $(0.5, 0.0, 5)$ (dashed curve). Resonances of $S$ in Fig. \ref{fig2}(b) are again
realized at about $T_0 \sim 9$ which coincides with the result 
in Fig. \ref{fig2}(a). This arises from the fact  
that an increase in the inverse of the Kramers rate 
with an increase of $\tau$ from $\tau=0.1$ to $\tau=5.0$ is approximately compensated 
by its decrease with an increase of $b$ from $b=0.1$ to $b=0.5$.
Figures \ref{fig2}(a) and \ref{fig2}(b) clearly show the resonant behavior 
of $S$ when a period of an external sinusoidal input $T_0$ is varied.
In the following, we will separately present 
simulation results for three cases: additive noise, multiplicative noise, 
and coexisting additive and multiplicative noises 
with fixed values of $T_0=10.0$ and $g=0.05$.

\begin{figure}
\begin{center}
\includegraphics[keepaspectratio=true,width=100mm]{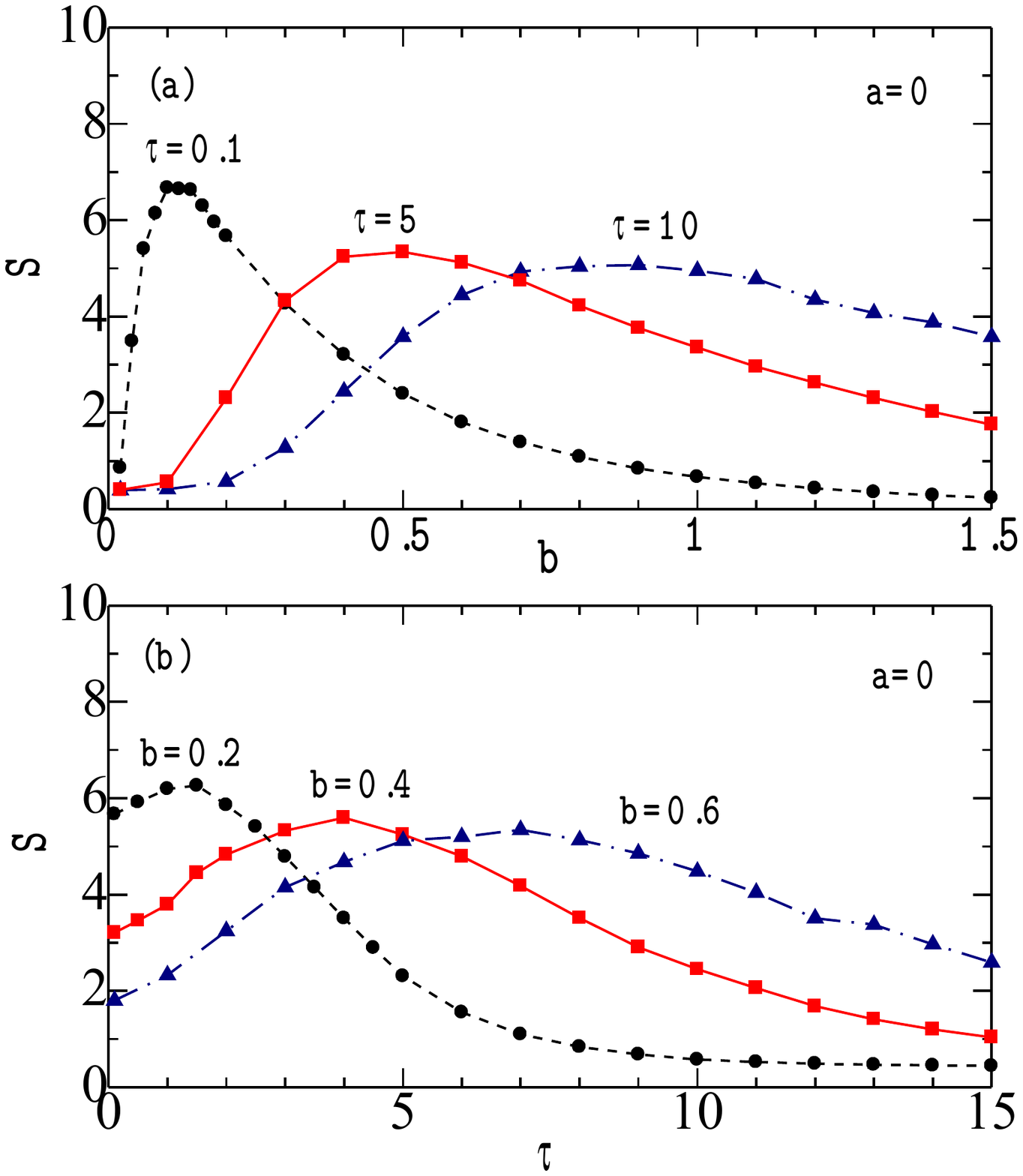}
\end{center}
\caption{
(Color online) 
(a) SPA $S$ for additive colored noise as a function of $b$
with $\tau=0.1$ (dashed curve), $\tau=5.0$ (solid curve)
and $\tau=10.0$ (chain curve).
(b) $S$ as a function of $\tau$ with $b=0.2$ (dashed curve), $b=0.4$ (solid curve)
and $b=0.6$ (chain curve) ($a=0.0$, $g=0.05$ and $T_0=10.0$).
}
\label{fig3}
\end{figure}

\begin{figure}
\begin{center}
\includegraphics[keepaspectratio=true,width=90mm]{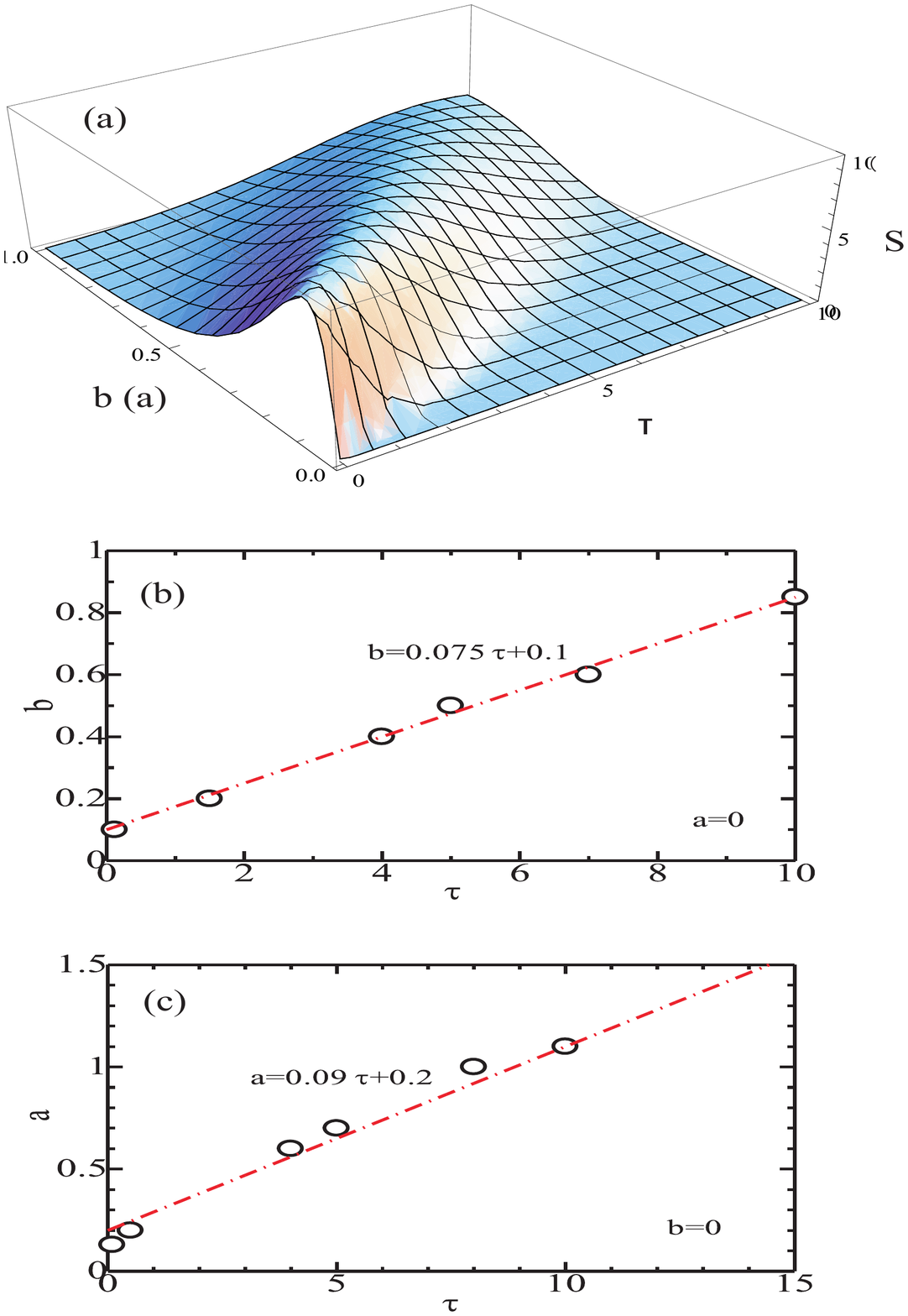}
\end{center}
\caption{
(Color online) 
(a) A schematic plot of $S$ as functions of $b$ and $\tau$ ($a$ and $\tau$)  
for additive (multiplicative) noise.
(b) Locations of $S_{max}$ for additive colored noise 
in the $b$-$\tau$ plane obtained by simulations (circles), 
the chain line being expressed by $b=0.075 \tau+0.1$.
(c) Locations of $S_{max}$ for multiplicative colored noise 
in the $a$-$\tau$ plane obtained by simulations (circles), 
the chain line being expressed by $a=0.09 \tau+0.2$.
}
\label{fig4}
\end{figure}

\subsubsection{Case of additive colored noise}
First we apply only additive colored noise ($a=0.0$) to the system.
Figure \ref{fig3}(a) shows SPA as a function of $b$
for $\tau=0.1$ (dashed curve),  $\tau=5.0$ (solid curve)
and $\tau=10.0$ (dashed curve).
With increasing $\tau$, the $b$ value where $S$ has the maximum is increased,
and the maximum value of $S$ ($S_{max}$) is gradually decreased with
the wider distribution of $S$.
Figure \ref{fig3}(b) shows the $\tau$ dependence of $S$ calculated  
for $b=0.2$ (dashed curve),  $b=0.4$ (solid curve)
and $b=0.6$ (dashed curve).
$S_{max}$ in Figs. \ref{fig3}(a) and \ref{fig3}(b) is observed 
at a larger $b$ for a larger $\tau$ and vice versa.

Figure \ref{fig4}(a) shows a schematic plot of $S$ as functions of $b$ and $\tau$, 
which is estimated from simulation results shown in Figs. \ref{fig3}(a) and \ref{fig3}(b). 
We note that with increasing $b$ or $\tau$,
a location of $S_{max}$ departs from the origin of $(b, \tau)=(0.0, 0.0)$
and the magnitude of $S_{max}$ is gradually decreased.
Circles in Fig. \ref{fig4}(b) denote locations of $S_{max}$ in the $b$-$\tau$ plane obtained
by simulations shown in Figs. \ref{fig3}(a) and (b).
$S_{max}$ nearly locates along the chain curve expressed by $b=0.075 \tau+0.1$.
Fig. \ref{fig4}(c) will be explained shortly.

\begin{figure}
\begin{center}
\includegraphics[keepaspectratio=true,width=100mm]{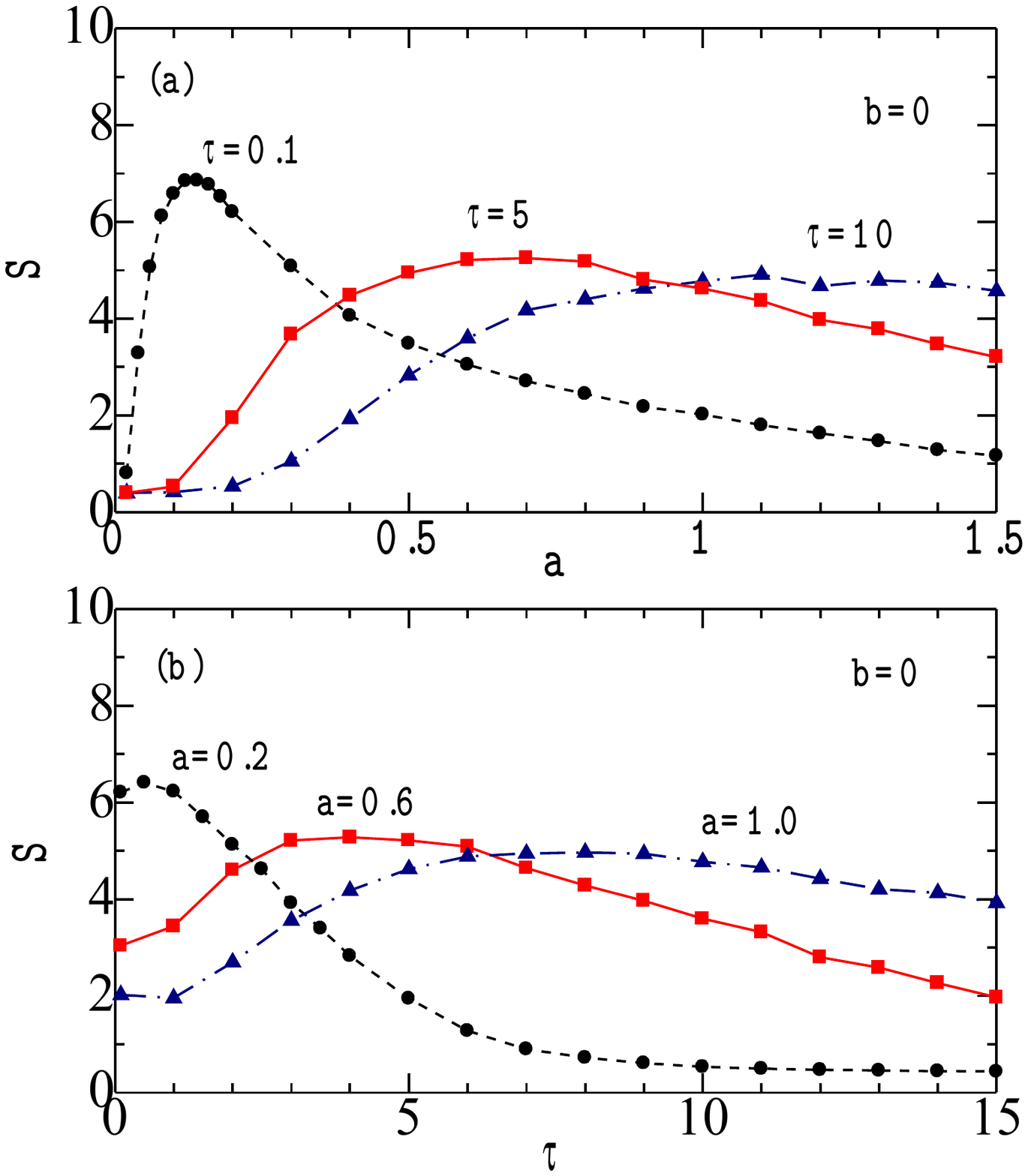}
\end{center}
\caption{
(Color online) 
(a) SPA $S$ for multiplicative colored noise as a function of $a$ 
with $\tau=0.1$ (dashed curve), $\tau=5.0$ (solid curve)
and $\tau=10.0$ (chain curve).
(b) $S$ as a function of $\tau$ with $a=0.2$ (dashed curve), $a=0.6$ (solid curve)
and $a=1.0$ (chain curve) ($b=0.0$, $g=0.05$ and $T_0=10.0$).
}
\label{fig5}
\end{figure}

\begin{figure}
\begin{center}
\includegraphics[keepaspectratio=true,width=100mm]{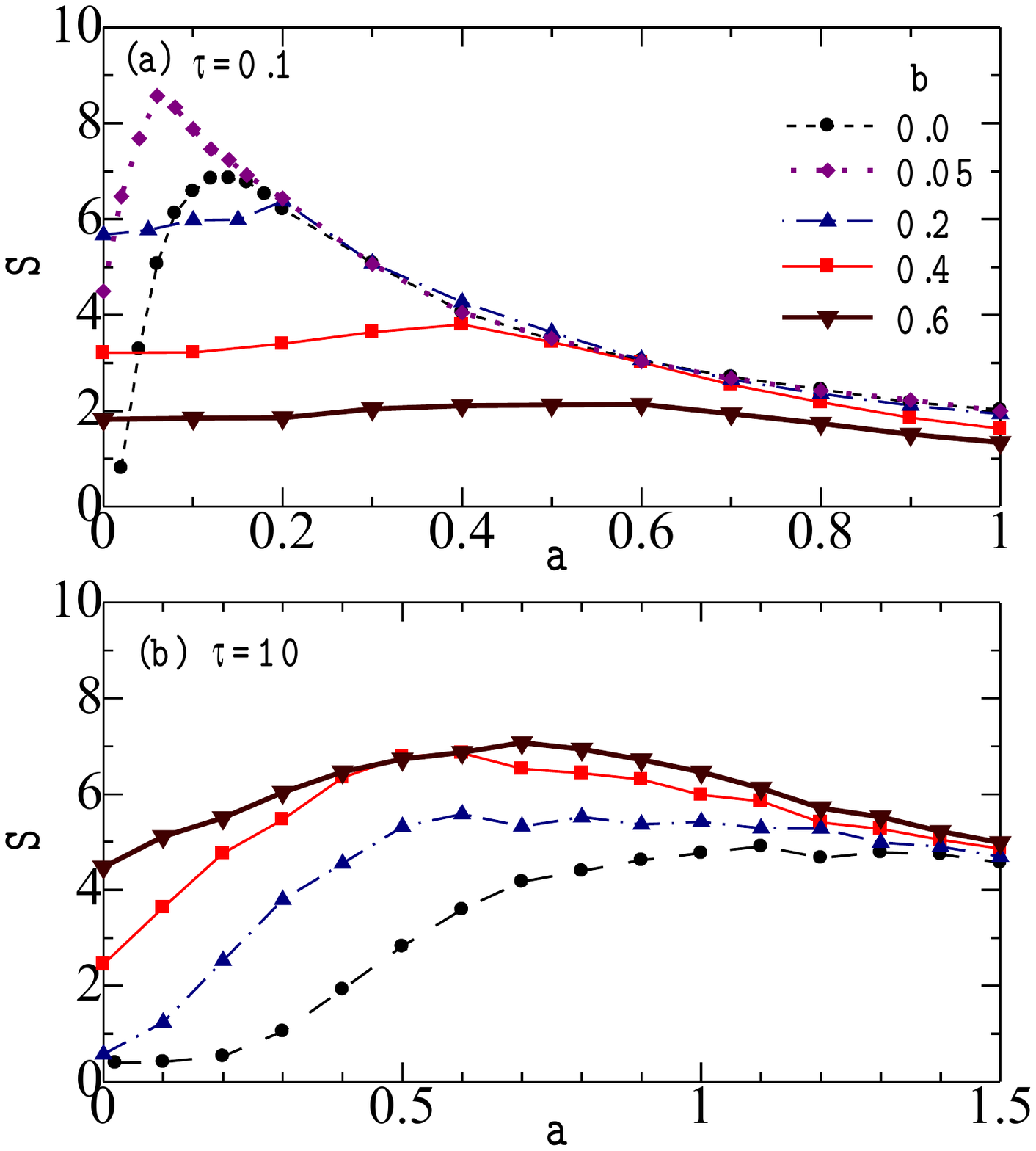}
\end{center}
\caption{
(Color online) 
SPA $S$ for coexisting additive and multiplicative colored noises
with (a) $\tau=0.1$ and (b) $\tau=10.0$:
$b=0.0$ (dashed curves), $b=0.05$ (dotted curves), 
$b=0.2$ (chain curves), $b=0.4$ (solid curves), 
and $b=0.6$ (bold solid curve) ($g=0.05$ and $T_0=10.0$).
}
\label{fig6}
\end{figure}

\begin{figure}
\begin{center}
\includegraphics[keepaspectratio=true,width=90mm]{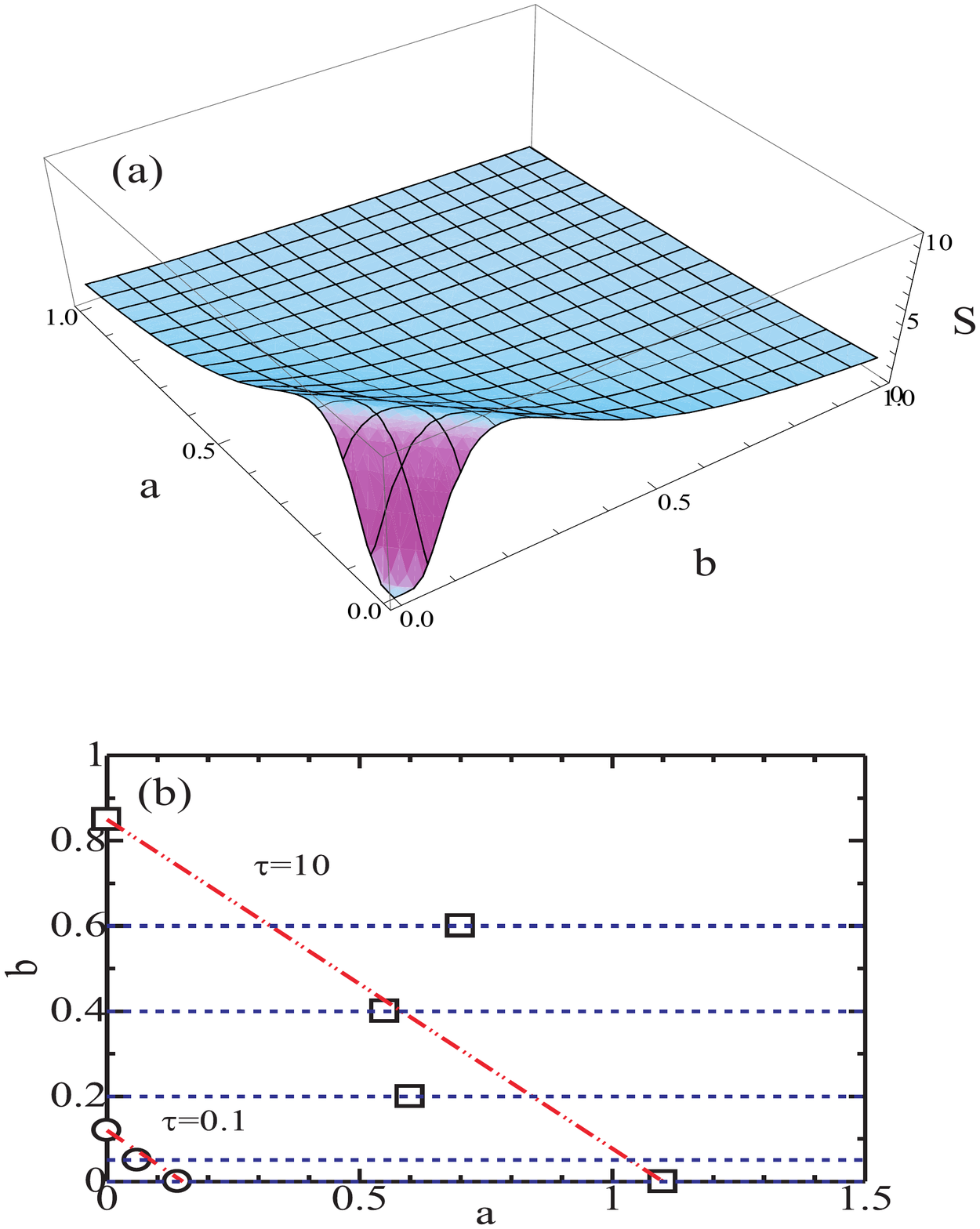}
\end{center}
\caption{
(Color online)
(a) A schematic plot of $S$ as functions of $a$ and $b$ for coexisting additive and 
multiplicative noises.
(b) Locations of $S_{max}$ in the $a$-$b$ plane: circles and squares denote
$S_{max}$ for $\tau=0.1$ and $\tau=10.0$, respectively, obtained by simulations,
chain and double-chain curves being shown for a guide of the eye. 
Along dashed lines, $a$ value in Fig. \ref{fig6}(a) or \ref{fig6}(b) is changed for a given $b$.
}
\label{fig7}
\end{figure}

\subsubsection{Case of multiplicative colored noise}
Figure \ref{fig5}(a) shows the $a$ dependence of $S$
when we apply only multiplicative colored noise with $\tau=0.1$ (dashed curve),  
$\tau=5.0$ (solid curve) and $\tau=10.0$ (dashed curve).
With increasing $\tau$, the position of $S_{max}$ moves to a larger $a$
and $S_{max}$ is gradually decreased.
Figure \ref{fig5}(b) shows the $\tau$ dependence of $S$ calculated  
for $a=0.2$ (dashed curve),  $a=0.6$ (solid curve)
and $b=1.0$ (dashed curve).

We expect from simulation results in Figs. \ref{fig5}(a) and \ref{fig5}(b)
that the $a$- and $\tau$-dependent PSA is given by
Fig. \ref{fig4}(a) where we read the $b$ axis as the $a$ axis. 
As in the case where additive colored noise is added,
a location of $S_{max}$ departs from the origin of $(a, \tau)=(0.0, 0.0)$
and $S_{max}$ is gradually decreased with increasing $a$ or $\tau$.
Circles in Fig. \ref{fig4}(c) denote locations of $S_{max}$ in the $a$-$\tau$ plane obtained
by simulations shown in Figs. \ref{fig5}(a) and (b).
$S_{max}$ nearly locates along the chain curve expressed by $a=0.09 \tau+0.2$.

\subsubsection{Case of coexisting additive and multiplicative colored noises}
Next we study the case where both additive and multiplicative colored noises are applied to the system.
Figure \ref{fig6}(a) shows the $a$ dependence of $S$ for $\tau=0.1$ with
$b=0.0$ (dashed curves), $b=0.05$ (dotted curve), $b=0.2$ (chain curves) and 
$b=0.4$ (solid curves).
The dashed curve for $b=0.0$ shows that $S$ has the maximum at $a \simeq 0.14$, 
as having been shown in Fig. \ref{fig5}(a).
With adding small additive noise of $b=0.05$, $S_{max}$ is increased
and its location moves to a smaller value of $a \simeq 0.05$.
With further increase of $b$ ($ \geq 0.2$), however, the magnitude of $S$
is gradually decreased and its maximum becomes obscure.
A similar $a$ dependence of $S$ for $\tau=10.0$ is plotted in Fig. \ref{fig6}(b) 
where $S$ is monotonously increased with increasing $b$.

Figure \ref{fig7}(a) expresses a schematic plot of $S$ as functions of $a$ and $b$
for the case of coexisting additive and multiplicative noises, 
which are estimated from results of simulations shown in Figs. \ref{fig6}(a) and \ref{fig6}(b).
$S$ has a small value at the origin of $(a, b)=(0.0, 0.0)$.
With departing from the origin by increases of $a$ and/or $b$, 
$S$ is first increased and then decreased after passing the maximum.
Circles and squares in Fig. \ref{fig7}(b) denote locations of $S_{max}$ 
in the $a$-$b$ plane obtained for $\tau=0.1$ and $\tau=10.0$, respectively,
from Figs. \ref{fig6}(a) and \ref{fig6}(b).
$S_{max}$ nearly locates along a chain (double-chain) curve for $\tau=0.1$ ($\tau=10.0$):
in the case of $\tau=10.0$ locations of $S_{max}$ are scattered 
because maximum of $S$ become obscure [Fig. \ref{fig6}(b)].

Dashed lines in Fig. \ref{fig7}(b) express traces along which $a$ value 
is changed for a given $b$ in Figs. \ref{fig6}(a) and \ref{fig6}(b).
With increasing $b$ in Fig. \ref{fig6}(a) for $\tau=0.1$, the magnitude of 
$a$-dependent $S$ is decreased
because a dashed line departs from the maximum expressed by a chain curve in Fig. \ref{fig7}(b),
except for $b=0.05$ for which the $a$-dependent $S$ is increased.
In contrast, with increasing $b$ in Fig. \ref{fig6}(b) for $\tau=10.0$, the magnitude of
the $a$-dependent $S$ is increased, particularly for $a < 0.5$, 
because a dashed line approaches the maximum expressed by a double-chain curve in Fig. \ref{fig7}(b).
These explain the difference of the $b$ dependences of $S$ in Figs. 7(a) and 7(b).

\section{Discussion}
\subsection{Underdamped Markovian approximation for additive noise}
We will examine the local approximation to the non-Markovian Langevin equation
given by Eq. (\ref{eq:A4}) for additive colored noise with $\phi'(x)=b$,
\begin{eqnarray}
\ddot{x} &=& - V'(x)- b^2 \int_{0}^t\gamma(t-t') \:\dot{x}(t') \:dt'
+ b \:\zeta(t)+ f(t),
\label{eq:N1} 
\end{eqnarray}
where the nonlocal kernel and OU colored noise are given by
\begin{eqnarray}
\langle \zeta(t)\zeta(t') \rangle &=& k_B T \:\gamma(t-t')
= \left( \frac{k_B T \gamma_0}{\tau}\right) \: e^{-\vert t-t' \vert/\tau}.
\label{eq:N2}
\end{eqnarray} 
As mentioned in Sec. II B, when employing the local limit given by Eq. (\ref{eq:B1}), 
\begin{eqnarray}
\left< \zeta(t) \zeta(t')\right> &=& k_B T \:\gamma(t-t') 
= 2 k_B T \:\gamma_0 \:\delta(t-t'),
\label{eq:N6}
\end{eqnarray}
we obtain the Markovian Langevin equation given by  
\begin{eqnarray}
\ddot{x} &=& -V'(x) - b^2 \gamma_0 \:\dot{x}
+\sqrt{2 k_B T} \:b  \:\xi(t) +f(t),
\label{eq:N7}
\end{eqnarray}
where $\xi(t)$ denotes white noise given by Eq. (\ref{eq:A15}).
If an alternative local approximation given by 
\begin{eqnarray}
\gamma(t-t') &=& 2 \gamma_0 \: \delta(t-t'),\;\;
\label{eq:N3} 
\langle \zeta(t)\zeta(t') \rangle 
=  \left( \frac{k_B T \gamma_0}{\tau}\right) \: e^{-\vert t-t' \vert/\tau},
\label{eq:N3b}
\end{eqnarray}
is adopted, Eq. (\ref{eq:N1}) reduces to the Markovian Langevin equation given by
\begin{eqnarray}
\ddot{x} &=& - V'(x)- b^2 \gamma_0 \:\dot{x} + b\:\zeta(t)+ f(t).
\label{eq:N4}
\end{eqnarray}
We should note that the memory kernel and colored noise given by Eq. (\ref{eq:N3b}) 
do not satisfy the FDR for $\tau \neq 0.0$, while those given by Eqs. (\ref{eq:N2}) 
and (\ref{eq:N6}) hold the FDR. 
The Markovian Langevin equation given by Eq. (\ref{eq:N4}) with OU colored noise 
[Eq. (\ref{eq:N3b})] has been employed for a study on SR in Ref. \cite{Gamma89}.
 
\begin{figure}
\begin{center}
\includegraphics[keepaspectratio=true,width=100mm]{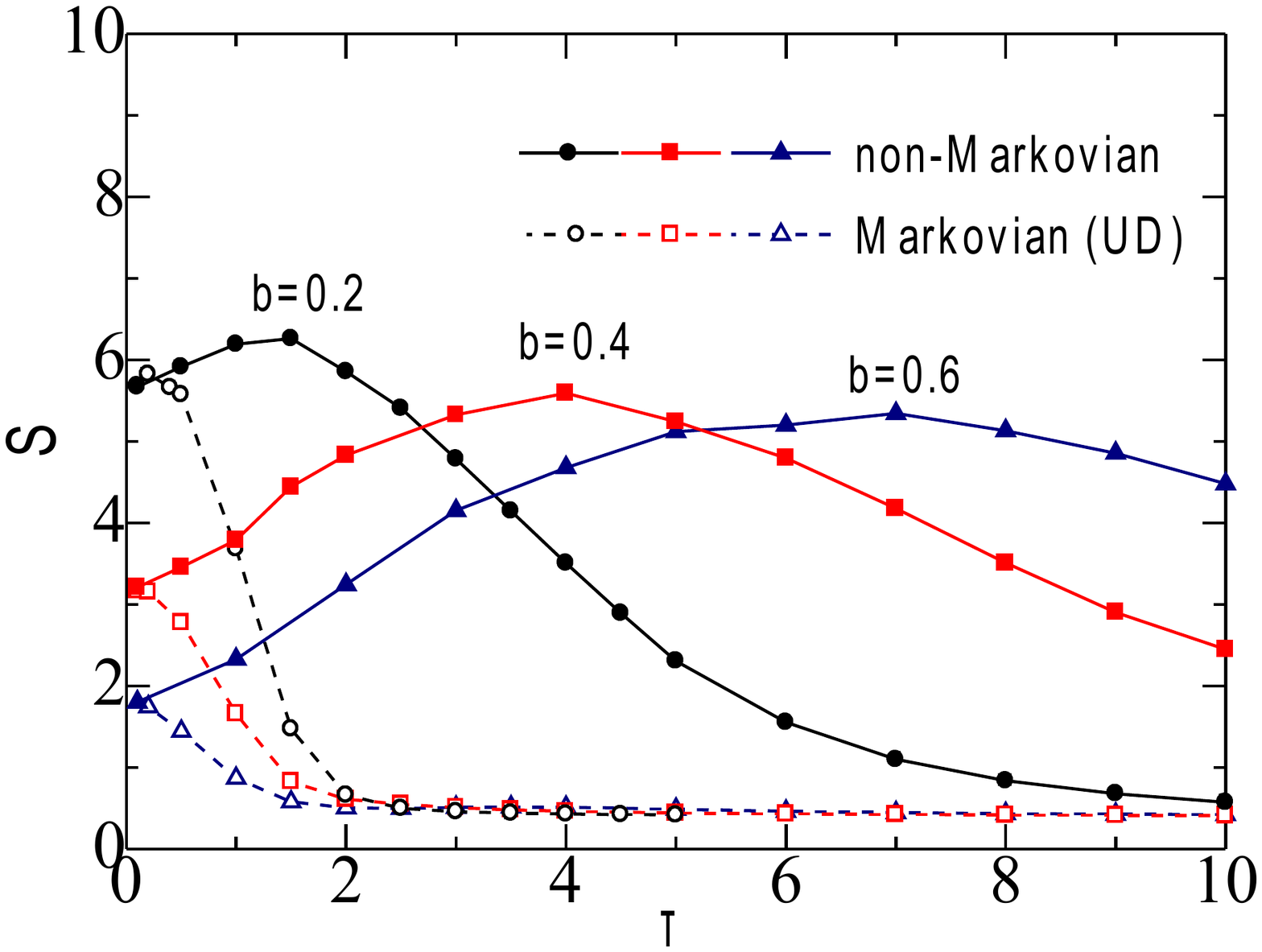}
\end{center}
\caption{
(Color online)
The $\tau$ dependence of $S$ for additive noise ($a=0.0$)
calculated by non-Markovian Langevin equation [Eq. (\ref{eq:N1})] (solid curves)
and underdamped (UD) Markovian approximation [Eq. (\ref{eq:N4})] (dashed curves) 
with $b=0.2$ (circles), $b=0.4$ (squares) and $b=0.6$ (triangles) 
($g=0.05$ and $T_0=10.0$).
}
\label{fig8}
\end{figure}

Solid curves in Fig. 8 show the $\tau$ dependence of $S$ calculated 
by the non-Markovian Langevin equation [Eq. (\ref{eq:N1})] 
for $b=0.2$ (circles), 0.4 (squares) and 0.6 (triangle)
with $k_B T=0.25$, $\gamma_0=1.0$, $g=0.05$ and $T_0=10.0$, 
which show the maxima as having been explained in Fig. 3(b).
Dashed curves express relevant plots of $S$ 
for (underdamped) Markovian Langevin equation given by Eq. (\ref{eq:N4})
with OU noise given [Eq. (\ref{eq:N3b})].
At $\tau \rightarrow 0.0$, results of the Markovian approximation agree 
with those of the non-Markovian Langevin equation as expected. 
With increasing $\tau$, $S$ in the Markovian approximation monotonously 
decreased as shown in Ref. \cite{Gamma89}, although for $b=0.2$, $S$ 
is slightly increased at $0.1 < \tau < 0.4$ but decreased at $\tau > 0.4$.
It is evident that results of the Markovian Langevin equation given by Eq. (\ref{eq:N4}) 
with the local approximation given by Eq. (\ref{eq:N3}) are quite different 
from those of the non-Markovian Langevin equation given by Eq. (\ref{eq:N1}) except for a very small $\tau$.

\begin{figure}
\begin{center}
\includegraphics[keepaspectratio=true,width=100mm]{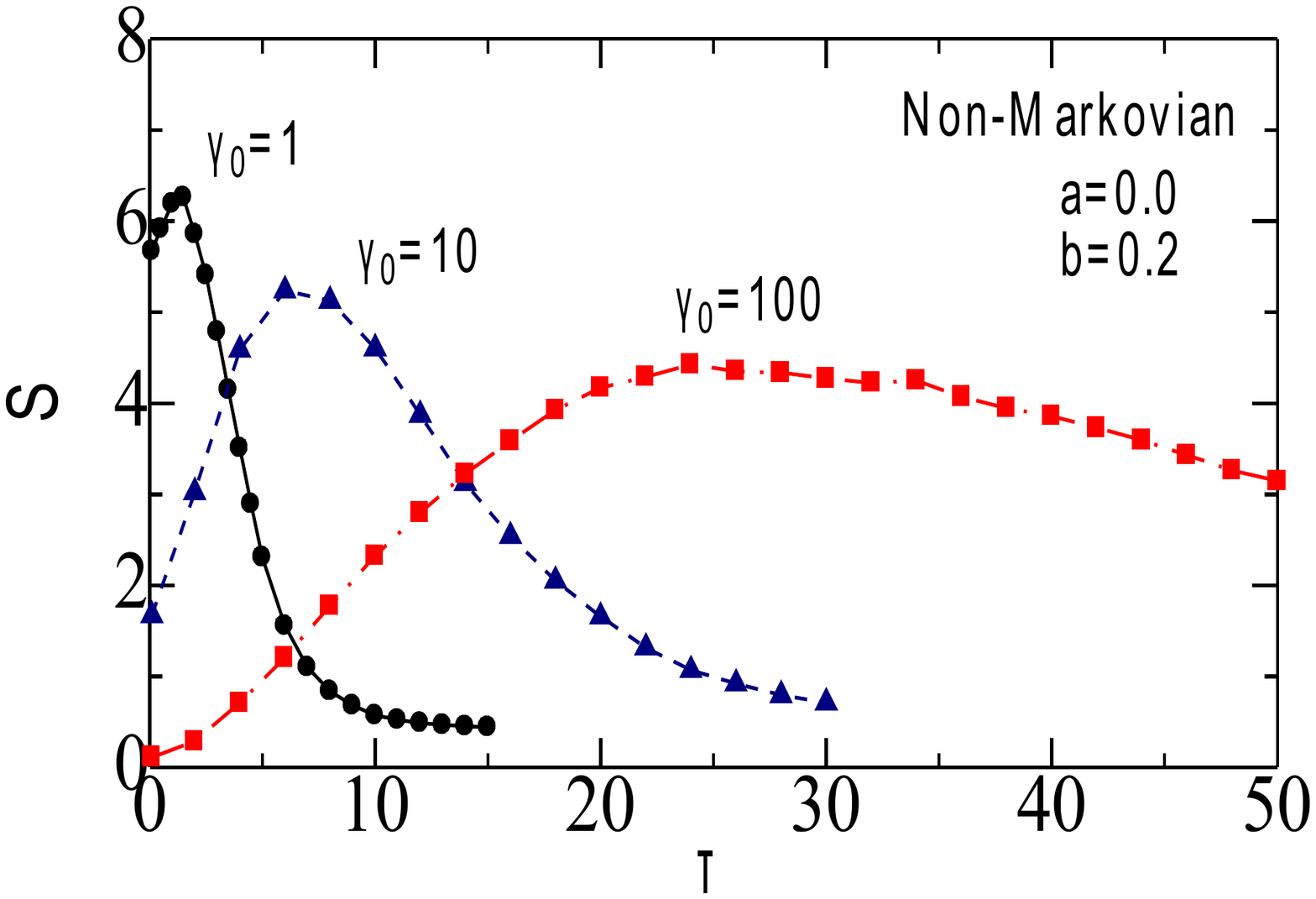}
\end{center}
\caption{
(Color online)
The $\tau$ dependence of $S$ for additive noise 
calculated by the non-Markovian Langevin equation [Eq. (\ref{eq:N1})] 
for $\gamma_0=1.0$ (solid curve), $\gamma_0=10.0$ (dashed curve) and $\gamma_0=100.0$ (chain curve)
with $a=0.0$, $b=0.2$, $g=0.05$ and $T_0=10.0$.
}
\label{fig9}
\end{figure}

\begin{figure}
\begin{center}
\includegraphics[keepaspectratio=true,width=100mm]{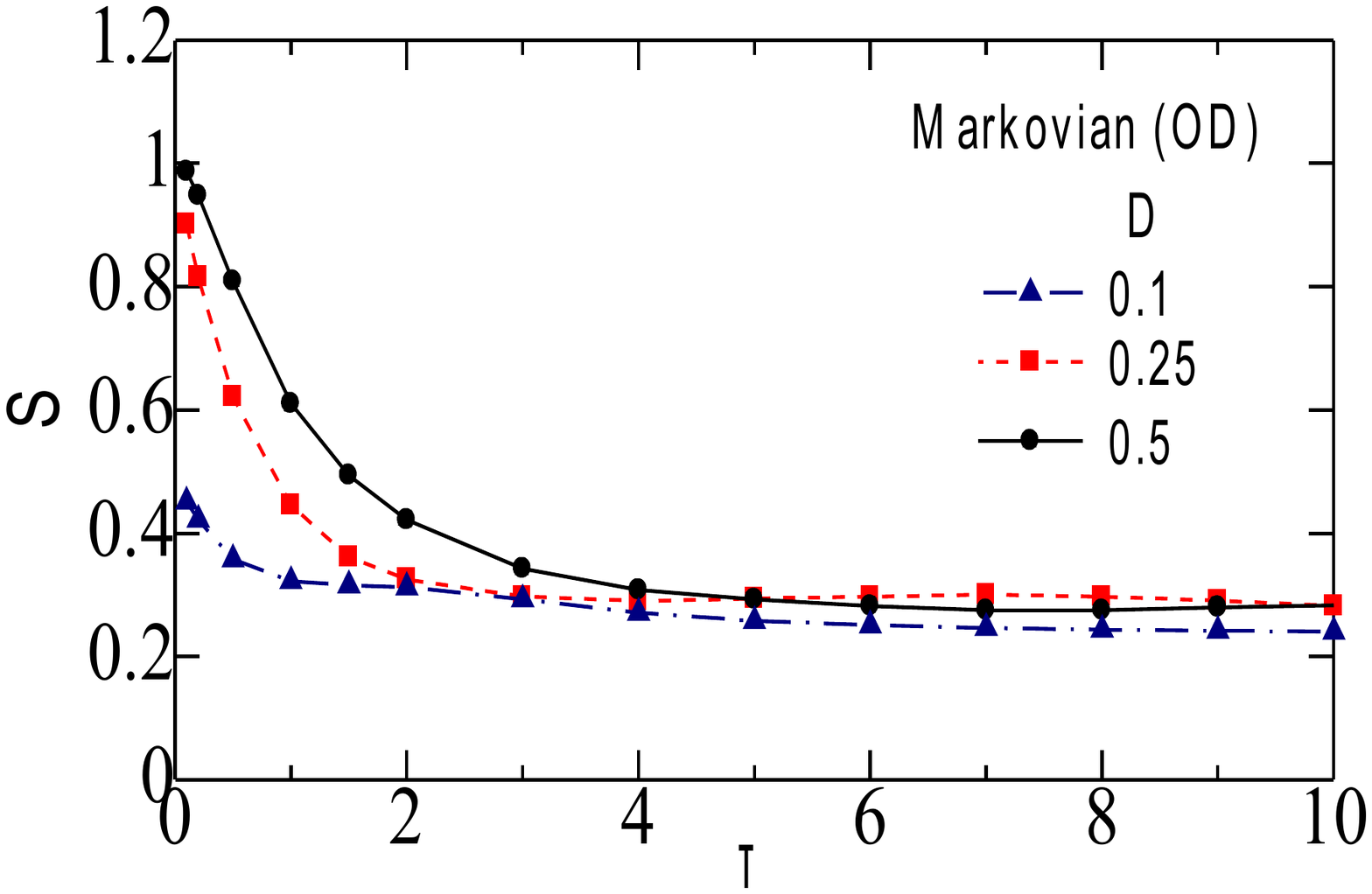}
\end{center}
\caption{
(Color online)
The $\tau$ dependence of $S$ for additive noise 
calculated by the overdamped (OD) Markovian approximation [Eq. (\ref{eq:N5})] 
with $D=0.10$ (chain curve), $D=0.25$ (dashed curve) and $D=0.50$ (solid curve) (see text).
}
\label{fig10}
\end{figure}

\subsection{Overdamped Markovian approximation for additive noise}

In order to gain an insight into SR of the overdamped limit of
the non-Markovian Langevin equation given by Eq. (\ref{eq:N1}) subjected to additive noise,
we have performed simulations, employing a large value of $\gamma_0$ in Eq. (\ref{eq:N2}).
Figure \ref{fig9} shows the $\tau$ dependence of $S$ for
$\gamma_0=1.0$ (solid curve), $10.0$ (dashed curve) and $100.0$ (chain curve)
with $a=0.0$, $b=0.2$, $g=0.05$ and $T_0=10.0$.
With a larger $\gamma_0$, $S(\tau)$ for $\tau \rightarrow 0$ becomes smaller
and the peak position for $S_{max}$ becomes larger. 
$S$ for all the cases investigated has the maximum.
Our calculation suggests that SR might be realized against $\tau$ even in the
overdamped limit of the non-Markovian Langevin equation.

In previous studies \cite{Hanggi84,Hanggi93}, the overdamped Langevin equation given by
\begin{eqnarray}
\dot{x} &=& - V'(x) + \zeta(t)+ f(t),
\label{eq:N5} 
\end{eqnarray}
with the OU noise correlation.
\begin{eqnarray}
\langle \zeta(t)\zeta(t') \rangle 
=  \left( \frac{D}{\tau}\right) \: e^{-\vert t-t' \vert/\tau},
\label{eq:N5b}
\end{eqnarray}
has been extensively studied within the UCNA. 
Equation (\ref{eq:N5}) may be obtainable from Eq. (\ref{eq:N4})
with setting $\ddot{x}=0$ and an appropriate change of variables.
  
Figure {\ref{fig10}} shows the $\tau$ dependence of $S$ calculated by simulations for
the overdamped Markovian Langevin equation given by Eq. (\ref{eq:N5}) with OU colored noise [Eq. (\ref{eq:N5b})]
for $D=0.10$ (chain curve), 0.25 (dashed curve) and 0.50 (solid curve).
With increasing $\tau$, all calculated results show monotonous decrease, 
in consistent with the result of the UCNA in Refs. \cite{Hanggi84,Hanggi93}. 
However, they exhibit no signs of the SR against $\tau$, which is in contrast with
the result in Fig. \ref{fig9} obtained by the non-Markovian Langevin equation
for a large $\gamma_0$. 

The discrepancy between results of non-Markovian Langevin equation [Eq. (\ref{eq:N1})] 
and Markovian Langevin equations [Eqs. (\ref{eq:N4}) and (\ref{eq:N5})] 
whose results are presented in Figs. 8, 9 and 10,
arises from the fact that the FDR is preserved in the former but in the latter.
We should note that a variation of $\tau$ in Eq. (\ref{eq:N2}) means 
simultaneous changes of the relaxation time of colored noise and of the correlation time 
of a memory kernel because they are mutually related by the FDR.
In contrast, a conventional local approach given by Eq. (\ref{eq:N3b}) or (\ref{eq:N5b})
includes a variation of $\tau$ only in the colored-noise relaxation time
with a vanishing memory.

Ref. \cite{Neiman96} has investigated the overdamped limit of Eq. (\ref{eq:N1}), 
taking account of the memory in a kernel given by 
\begin{eqnarray}
k_B T\gamma(t-t') &=& \langle \xi(t) \xi(t') \rangle  
= 2 \gamma_0 \:\delta(t-t')
+ \left( \frac{\gamma_1}{\tau} \right) \:e^{-\vert t-t'\vert/\tau},
\label{eq:L8b}
\end{eqnarray}
where $\gamma_0$ and $\gamma_1$ express magnitudes of
local and non-local  dissipations, respectively.
Equations  (\ref{eq:N1}) and (\ref{eq:L8b}) yield the over-damped Markovian Langevin equation 
given by
\begin{eqnarray}
\gamma_0 \dot{x} &=& -V'(x)-\frac{\gamma_1 x}{\tau}+\frac{\gamma_1 x(0) e^{-t/\tau}}{\tau}
+\left( \frac{\gamma_1}{\tau^2} \right) \int_0^t e^{-(t-t')/\tau} x(t')\:dt'+\xi(t)+f(t).
\label{eq:L8c}
\end{eqnarray}
It has been shown that when $\tau$ is increased, the SR is first
suppressed for small $\tau$ but enhanced for large $\tau$ 
with the minimum at intermediate $\tau$ \cite{Neiman96}.
Although the FDR is held in Eq. (\ref{eq:L8b}), its expression cannot be justified
from a microscopic point of view. 

\subsection{Overdamped Markovian approximation for additive and multiplicative noises}
For additive and multiplicative noises with $\phi'(x)=ax + b$,
Eqs. (\ref{eq:D2}), (\ref{eq:D3}) and (\ref{eq:D5}) in Sec. II C 
lead to the overdamped Markovian Langevin equation given by 
\begin{eqnarray}
\dot{x} &=& - \frac{V'(x)}{\gamma_0 (a x + b)^2}
-\frac{ k_B T a}{\gamma_0 (a x + b)^3} 
+\sqrt{ \frac{2 k_B T}{\gamma_0 \: (a x + b)^2} } \:\xi(t),
\label{eq:D7}
\end{eqnarray}
where $\xi(t)$ is white noise given by Eq. (\ref{eq:A15}).
Equation (\ref{eq:D7}) is, however, quite different from a widely-adopted phenomenological
Langevin model given by 
\begin{eqnarray}
\dot{x} &=& - V'(x)+ \bar{\xi}(t)+  x \:\bar{\eta}(t),
\label{eq:D8}
\end{eqnarray}
with correlations,
\begin{eqnarray}
\langle \bar{\xi}(t) \bar{\xi}(t') \rangle &=& 2D \:\delta(t-t'),\;\;
\langle \bar{\eta}(t) \bar{\eta}(t') \rangle = 2D_m \;\delta(t-t'), \nonumber \\
\langle \bar{\xi}(t) \bar{\eta}(t')\rangle &=&  0,
\label{eq:D8b}
\end{eqnarray}
where $D$ and $D_m$ stand for magnitudes of additive and multiplicative noises, respectively.
The stationary PDF obtained from the PFE for Eqs. (\ref{eq:D8}) and (\ref{eq:D8b})
in the Stratonovich sense is given by
\begin{eqnarray}
\ln P(x) &=& - \int \frac{V'(x)}{(D+D_m x^2)}\;dx 
- \left( \frac{1}{2}\right) \ln(D+D_m x^2).
\label{eq:D9}
\end{eqnarray}
For the parabolic potential of $V(x)=x^2/2$, Eq. (\ref{eq:D9}) yields Gaussian or non-Gaussian PDF,
depending on $D$ and $D_m$ \cite{Sakaguchi01,Anteneodo03,Hasegawa07}.
For the bistable potential of $V(x)=x^4/4-x^2/2$, Eq. (\ref{eq:D9}) leads to
\begin{eqnarray}
P(x) &\propto& (D+D_m x^2)^{-(D_m^2-D_m-D)/2 D_m^2} \;e^{-x^2/2 D_m}
\hspace{0.5cm}\mbox{for $D \neq 0.0$, $D_m\neq 0.0$}, \\
&\propto& e^{-(x^4/4-x^2/2)/D}
\hspace{4.5cm}\mbox{for $D_m=0.0$}, \\
&\propto& \vert x \vert^{1/D_m-1} \;e^{-x^2/2D_m}
\hspace{4cm}\mbox{for $D=0.0$}.
\end{eqnarray}
Thus the stationary PDF of Eq. (\ref{eq:D9}) depends on magnitudes of additive and
multiplicative noises, in contrast with that of Eq. (\ref{eq:D6}) 
which is independent of them.

Studies on SR have been made in Refs. \cite{Li95,Fu99,Jia00,Jia01,Luo03},
by using the Langevin equation given by Eq. (\ref{eq:D8}) but with
generalized correlations,
\begin{eqnarray}
\langle \bar{\xi}(t) \bar{\xi}(t') \rangle &=& 2 D \:\delta(t-t'), \;\;
\langle \bar{\eta}(t) \bar{\eta}(t') \rangle = (D_m/\tau) \;e^{-\vert t-t'\vert /\tau}, 
\nonumber \\
\langle \bar{\xi}(t) \bar{\eta}(t')\rangle &=&  2 \lambda \sqrt{D D_m}\;\delta(t-t'),
\label{eq:L2b}
\end{eqnarray}
where $\lambda$ is the coupling strength between additive and multiplicative noises.
SNR calculated with the use of the UCNA and the functional integral method
shows a complicated behavior as functions of $D$, $D_m$ and $\lambda$. 
However, the phenomenological Langevin equation given by Eq. (\ref{eq:D8}) with Eq. (\ref{eq:L2b}) 
has no microscopic bases and does not agree with microscopically 
obtained expression given by Eq. (\ref{eq:D7}).

\section{Concluding remarks}
Phenomenological Langevin equations mentioned in the preceding section have following deficits:

\noindent
(a) the FDR does not hold between a local dissipation and nonlocal colored noise in Eq. (\ref{eq:N3b}) 
for the Langevin equation given by Eqs. (\ref{eq:N4}) or (\ref{eq:N5}) except for $\tau = 0$, 

\noindent
(b) the nonlocal kernel and colored noise in Eq. (\ref{eq:L8b}) for the Langevin equation 
given by Eq. (\ref{eq:L8c}) have no microscopic bases although they hold the FDR, and

\noindent
(c) the FDR in the Langevin equation given by Eq. (\ref{eq:D8}) with additive and multiple noises 
whose correlations are given by Eq. (\ref{eq:D8b}) or (\ref{eq:L2b}) is not definite.

\noindent
The non-Markovian Langevin equation given by Eq. (\ref{eq:A4}) derived in this study from 
the generalized CL model
is free from the deficits (a)-(c) in phenomenological Langevin models.
We have studied the open bistable system by simulations, obtaining the following results:

\noindent
(i) calculated marginal PDFs for $x$ and $p$ are given by
$P(x) \propto e^{-\beta V(x)}$ and $P(p) \propto e^{-\beta p^2/2}$, respectively,
independently of noise parameters of $a$, $b$ and $\tau$ (Fig. 1) although SPA depends on them,  

\noindent
(ii) SPA exhibits SR for a variation of $\tau$ 
[Figs. \ref{fig3}(b) and \ref{fig5}(b)] besides those of $a$ and $b$ 
[Figs. \ref{fig3}(a) and \ref{fig5}(a)], 

\noindent
(iii) SPA for coexisting additive and multiplicative noises shows unimodal
but bimodal structures as functions of $a$, $b$ and/or $\tau$
[Figs. \ref{fig4}(a) and \ref{fig7}(a)], and

\noindent
(iv) the local approximation [Eq.(\ref{eq:N3})] to non-Markovian Langevin equation
cannot provide a reliable description except for a vanishingly small $\tau$ (Fig. \ref{fig8}).

\noindent
Items (i)-(iii) are in contrast with the results obtained by phenomenological Langevin models
\cite{Hanggi84,Hanggi93,Gamma89,Neiman96,Li95,Fu99,Jia00,Jia01,Luo03} 
where the FDR is indefinite or not preserved. 
In particular, the $\tau$ dependence of SR in the item (ii) is different even qualitatively
from previous results: a monotonous decrease of SNR
with $\tau$ \cite{Hanggi84,Hanggi93,Gamma89} and the minimum of SNR 
at intermediate $\tau$ \cite{Neiman96}.  
It would be interesting to apply the present approach to a study of relevant phenomena
in bistable systems such as the first passage time and resonant activation.


\begin{acknowledgments}
This work is partly supported by
a Grant-in-Aid for Scientific Research from 
Ministry of Education, Culture, Sports, Science and Technology of Japan.  
\end{acknowledgments}


\end{document}